\pdfoutput=1

\documentclass[natbib]{svjour3}                     
\smartqed  
\usepackage{aps-bibstyle}  
\usepackage[pdftex]{graphicx}

\journalname{Space Sience Review}
\def\arcdeg{\hbox{$^\circ$}}

\def\arcsec{\hbox{$^{\prime\prime}$}}
\def\sun{\hbox{$\odot$}}
\def\gtrsim{\mathrel{\hbox{\rlap{\hbox{\lower4pt\hbox{$\sim$}}}\hbox{$>$}}}}
\def\lesssim{\mathrel{\hbox{\rlap{\hbox{\lower4pt\hbox{$\sim$}}}\hbox{$<$}}}}

\begin{document}

\title{What Are Special About Ground-Level Events?}

\subtitle{Flares, CMEs, Active Regions And Magnetic Field Connection}


\author{N.V. Nitta \and Y. Liu \and M.L. DeRosa \and R.W. Nightingale}

\authorrunning{N.V. Nitta et al.} 

\institute{N.V. Nitta, M.L. DeRosa, R.W. Nightingale\at
              Lockheed Martin Solar and Astrophysics Laboratory, Dept/ADBS, B/252, 3251 Hanover Street, Palo Alto, CA 94304, USA \\
              Tel.: +1 650-354-5458\\
              Fax: +1 650-424-3994\\
              \email{nitta@lmsal.com}           
           \and
           Y. Liu \at
              Stanford University, Stanford, 94305, USA
}

\date{Received: date / Accepted: date}

\maketitle

\begin{abstract}
Ground level events (GLEs) occupy the high-energy end of
gradual solar energetic particle (SEP) events.  They are
associated with coronal mass ejections (CMEs) and solar flares, but we
still do not clearly understand the special conditions that produce
these rare events.  During Solar Cycle 23, a total of 16 GLEs were
registered, using ground-based neutron monitor data.  
We first ask if these GLEs are clearly distinguishable from other SEP
events observed from space.  Setting aside possible difficulties in 
identifying all GLEs consistently, 
we then try to find observables which may unmistakably
isolate these GLEs by studying the basic properties of the associated
eruptions and the active regions (ARs) that produced them.  It is
found that neither the magnitudes of the CMEs and flares nor the
complexities of the ARs give sufficient conditions for
GLEs.  It is possible to find CMEs, flares or ARs that are
not associated with GLEs but that have more extreme properties than
those associated with GLEs.  
We also try to evaluate the importance of magnetic field connection of
the AR with Earth on the detection of GLEs and their onset times.
Using the potential field source surface (PFSS) model, a half
of the GLEs are found to be well-connected.  However, 
the GLE onset time with respect to the onset of the associated 
flare and CME 
does not strongly depend on how well-connected the AR is.  
The GLE onset behavior may be largely determined by 
when and where the CME-driven shock develops.  
We could not relate the shocks responsible for the onsets of past GLEs
with features in solar images, but 
the combined data from the Solar
TErrestrial RElations Observatory (STEREO) and the Solar Dynamics
Observatory (SDO) have the potential to change this for
GLEs that may occur in the rising phase of Solar Cycle 24.

\end{abstract}

\section{Introduction}
\label{intro}
Ground-level events (GLEs) are a special class of solar energetic
particle (SEP) events, in which ions are accelerated to relativistic
energies and cause ground-level effects. As of the second half of 2011, 
there have been only 70 GLEs on
record since 1942, including 16 during Solar Cycle 23
\citep{Gopal10,Gopal11a}.  Though detailed
case studies have been conducted on a number of individual GLEs,
we still do not fully understand the conditions and processes that
are responsible for these extreme SEP events. 

With a high $\sim$10~MeV proton flux, GLEs are usually identified with
so-called gradual (as opposed to impulsive) SEP events,
which are accelerated at shocks driven by coronal mass
ejections (CMEs) \citep[e.g.,][]{Reames99,Kahler11b}. 
Another view, however, is that some
GLEs come directly from solar flares, which may also accelerate particles to
high energies in magnetic reconnection,
primarily because their temporal
variations mimic those of impulsive flares 
\citep{Grechnev08, McCracken08}. 
\citet{Aschwanden12} also argued that, in five of the 13 GLEs he
studied, the
particles are released during the impulsive phase of the associated flare.
However, it could be problematic to advocate the flare origin of GLEs 
on the basis of temporal variations alone,
because the flare impulsive phase tends to be
contemporaneous with 
the CME rapid acceleration phase \citep{Zhang01,Temmer10}, which
may be the likeliest time for the formation of a shock wave.
In addition, direct contributions of flare-accelerated particles in
GLEs may be suggested by the SEP compositions and charge states
similar to those of impulsive SEP events 
\citep[see, e.g.,][for GLEs~55\,--\,57]{Cohen99}.

GLEs tend to be associated with fast CMEs and intense
flares \citep[e.g.,][and Section~2]{Gopal10,Gopal11a}.
Apart from the magnitudes of the associated CMEs and flares, there seem to be
at least two more factors that may contribute to large SEP events in
near-Earth space including GLEs.  
One is the presence of prior CMEs.  \citet{Gopal04}
showed statistically that large SEP events tend to be
associated with CMEs preceded by another CME
from the same region within a 24 hour time window.  
Similar conclusions were reached by \citet{Kahler05}.
The preceding CME may perform pre-conditioning in two ways: to provide
seed particles to be re-accelerated at the shock driven by the main CME 
\citep{Kahler01,Cliver83,Cliver06}, 
and to cause higher level of turbulence in the
upstream region of the main CME \citep{Li05_icrc}.
Recently \cite{Li11} have further developed the concept of double CMEs
occurring close in time that may lead to enhanced particle acceleration.

Another factor is magnetic field connection of the acceleration region
to Earth.  The source regions of gradual SEP events, usually active
regions (ARs), are distributed in 
much broader longitudes than those of impulsive SEP events \citep{Reames99},
presumably reflecting larger extensions of the CME-driven shocks than
the reconnection regions above flares.  The broad longitudinal
distribution of source ARs may discount the importance of their
magnetic field connection to the observer.  
However, the time profiles of gradual SEP
events depend on the source longitudes.  For example, those from
the western source ARs tend to rise more quickly to the peak than 
those from the eastern source regions \citep{Cane88,Reames99,Cane03}.  
This indicates that magnetic field connection of the source region
and surrounding 
area 
may play a role in the SEP onset and 
peak times and possibly in other
SEP properties, given that the western hemisphere is more likely connected
to Earth in normal interplanetary conditions.  Note that the very
intense GLE on 20 January 2005 was associated with a flare around the
longitude of W60.

Therefore the origin of large SEP events including GLEs may not be 
understood properly from the
associated discrete solar events alone.  Nevertheless, it is still
meaningful to ask if there are special properties of the flares and
CMEs that are associated with GLEs, as well as of the ARs
that produce them.  In particular, the properties of ARs have
only rarely been discussed with regard to large SEP events since space-borne
solar data became routinely available in the 1990s 
\citep[e.g.,][]{Nitta03_icrc,Gopal05_iau,Kahler11b}, and so we choose them as
one of the central themes here.  

This article is organized as follows.
In Section~2, we discuss possible problems in 
defining and characterizing GLEs.
We consider this to be 
important to set the scope of the following sections.
In Section 3, we review the magnitudes of the flares and
CMEs that were associated with GLEs in Solar Cycle 23.  
Section 4 compares some basic properties of the ARs that produced
GLEs with those of (a) ARs that produced SEP events but not
GLEs, and of (b) complex ARs in general.  In Section 5, 
we discuss magnetic field connection of the source ARs and surrounding
regions with Earth.  We describe in Section 6 possible signatures of 
shocks that may be relevant to the onset of GLE events.
We discuss and summarize this study in Section~7.  Many topics,
although possibly important, have to be excluded, such as cross field
diffusion and the effect of prior CMEs and seed particles.

\section{Problems in defining and characterizing GLEs}
\label{sec:2}

\begin{table}
\caption{GLE events and associated flares and CMEs (adopted from Gopalswamy et
  al. 2010)} 
\begin{tabular}{rrrrrrrrrr}
\hline\noalign{\smallskip} 
\multicolumn{4}{c}{GLE} & & \multicolumn{2}{c}{Flare} & & \multicolumn{2}{c}{CME} \\
\cline{1-4} \cline{6-7} \cline{9-10} \noalign{\smallskip}
& \multicolumn{2}{c}{Onset} & Max & & \multicolumn{1}{c}{GOES} & & & POS & Width \\
ID & \multicolumn{1}{c}{Date} & \multicolumn{1}{c}{Time$^{a}$} &
\multicolumn{1}{c}{Int (\%)$^{a}$} & & \multicolumn{1}{c}{Class} &
\multicolumn{1}{c}{Location} 
& & \multicolumn{1}{c}{Speed (km/s)} & \multicolumn{1}{c}{(degs)} \\
\noalign{\smallskip}\hline\noalign{\smallskip}
55 & 1997/11/06 & 12:10 & 11.3 & & X9.4  & S18W63 & & 1556 & 360   \\
56 & 1998/05/02 & 13:55 &  6.8 & & X1.1  & S15W15 & & 938 & 360   \\
57 & 1998/05/06 & 08:25 &  4.2 & & X2.7  & S11W66 & & 1099 & 190   \\
58 & 1998/08/24 & 22:50 &  3.3 & & X1.0  & N35E09 & & --$^{b}$ & --$^{b}$   \\
59 & 2000/07/14 & 10:30 & 29.3 & & X5.7  & N22W07 & & 1674 & 360 \\
60 & 2001/04/15 & 14:00 & 56.7 & & X14  & S20W85 & & 1199 & 167 \\
61 & 2001/04/18 & 02:35 & 13.8 & & C2.2 & S20W116 & & 2465 & 360 \\
62 & 2001/11/04 & 17:00 &  3.3 & & X1.0  & N06W18 & & 1810 & 360 \\
63 & 2001/12/26 & 05:30 &  7.2 & & M7.1  & N05W54 & & 1446 & $>$212 \\
64 & 2002/08/24 & 01:18 &  5.1 & & X3.1  & S02W81 & & 1913 & 360 \\
65 & 2003/10/28 & 11:22 & 12.4 & & X17   & S18E18 & & 2459 & 360 \\
66 & 2003/10/29 & 21:30 &  8.1 & & X10   & S18W04 & & 2029 & 360 \\
67 & 2003/11/02 & 17:30 &  7.0 & & X8.3  & S18W57 & & 2598 & 360 \\
68 & 2005/01/17 & 09:55 &  3.0 & & X3.8  & N14W25 & & 2547 & 360 \\
69 & 2005/01/20 & 06:51 &277.3 & & X7.1  & N14W61 & & 3242$^{c}$ & 360 \\
70 & 2006/12/13 & 02:45 & 92.3 & & X3.4  & S06W23 & & 1774 & 360 \\

\noalign{\smallskip}\hline
\noalign{\bigskip}

\end{tabular}

\noindent
a. According to the Oulu Neutron Monitor. 

\noindent
b. No SOHO LASCO data. 

\noindent
c. From Gopalswamy et al. (2010).  There are different estimates
(see Grechnev et al. 2008).

\end{table}

GLEs consist of relativistic ions.  Like galactic cosmic-rays, they 
penetrate into Earth's atmosphere and
produce secondary particles.  The 16 GLEs registered during
Solar Cycle 23 are summarized in Table~1.  Whether a given SEP event is a GLE
depends on the detection of the secondary particles usually by neutron
monitors (NMs) at cosmic ray stations. For a given GLE, different NMs
show different time profiles \citep[e.g.,][]{Bieber02, McCracken08, Moraal11},
because of different asymptotic look angles and rigidities of the NMs
combined with
anisotropy of the particles.  

\footnotetext[1]{\url{http://cosmicrays.oulu.fi}}

Therefore, even though 
the onset time and maximum intensity in Table~1 (columns 3 and 4) 
are based on data from the Oulu Cosmic Ray Station 
(with a $\approx$0.8 GV cutoff
rigidity)\footnotemark  as was the case with \cite{Gopal10}, 
we acknowledge that NM data with a wide range of location and
cutoff rigidity are needed to characterize a GLE including 
the proton spectrum
\citep[e.g.][]{Tylka09,Matthia09}.  
Even for simpler quantities such as the onset time and peak flux,
it is important to 
find a NM whose asymptotic look angle is nearly
aligned with the interplanetary magnetic field.  Such a NM with a low
($\lesssim$1~GV) cutoff rigidity is expected
to give the earliest onset and highest increase in count rate.  For
example, \cite{Shea11} give these NMs for all the GLEs in Solar
Cycle 23. The Oulu NM is chosen for seven of the 16 GLEs.
Different stations are chosen for the remaining GLEs, and all of them
have the rigidity of 1~GV or less.  

It is of interest to compare the time
profiles of the NM count rates with those of the proton fluxes
directly measured from space.
The highest-energy measurement of protons from space is achieved by 
the High Energy Proton and Alpha
Detector \citep[HEPAD;][]{Onsager96} on the Geostationary
Operations Environmental Satellite ({\it GOES}), which
provides differential fluxes in three channels between 350 MeV and
700 MeV, and integral flux above 700 MeV.  Figure~1 gives such a comparison
of the HEPAD P9 (420\,--\,510~MeV) data with the NM data that recorded 
the earliest onset and/or the highest peak count. 
The HEPAD data in five-minute
average are available at the National Geophysics Data Center\footnotemark.  
Note that the 1~GV rigidity cutoff
corresponds to the proton energy of $\approx$430~MeV, so we expect
protons in the P9 channel to overlap with those detected by
$\leq$1~GV NMs.

In Figure~1, the time profiles are plotted in a
time range between 10 minutes before and 110 minutes after the onset of
the associated flare.  Here we are interested only in 
the general temporal behavior of GLEs at their onsets, 
so we do not go through the
procedures necessary to correct proton spectra from the HEPAD data 
\citep[see][]{Smart99}.  Differences among the HEPAD data from
different {\it GOES} satellites are also beyond the scope of these
comparisons; data from the ``primary'' satellite should be official.

\footnotetext{\url{http://goes.ngdc.noaa.gov/data/avg/}}

\begin{figure*}
\includegraphics[width=1.0\textwidth]{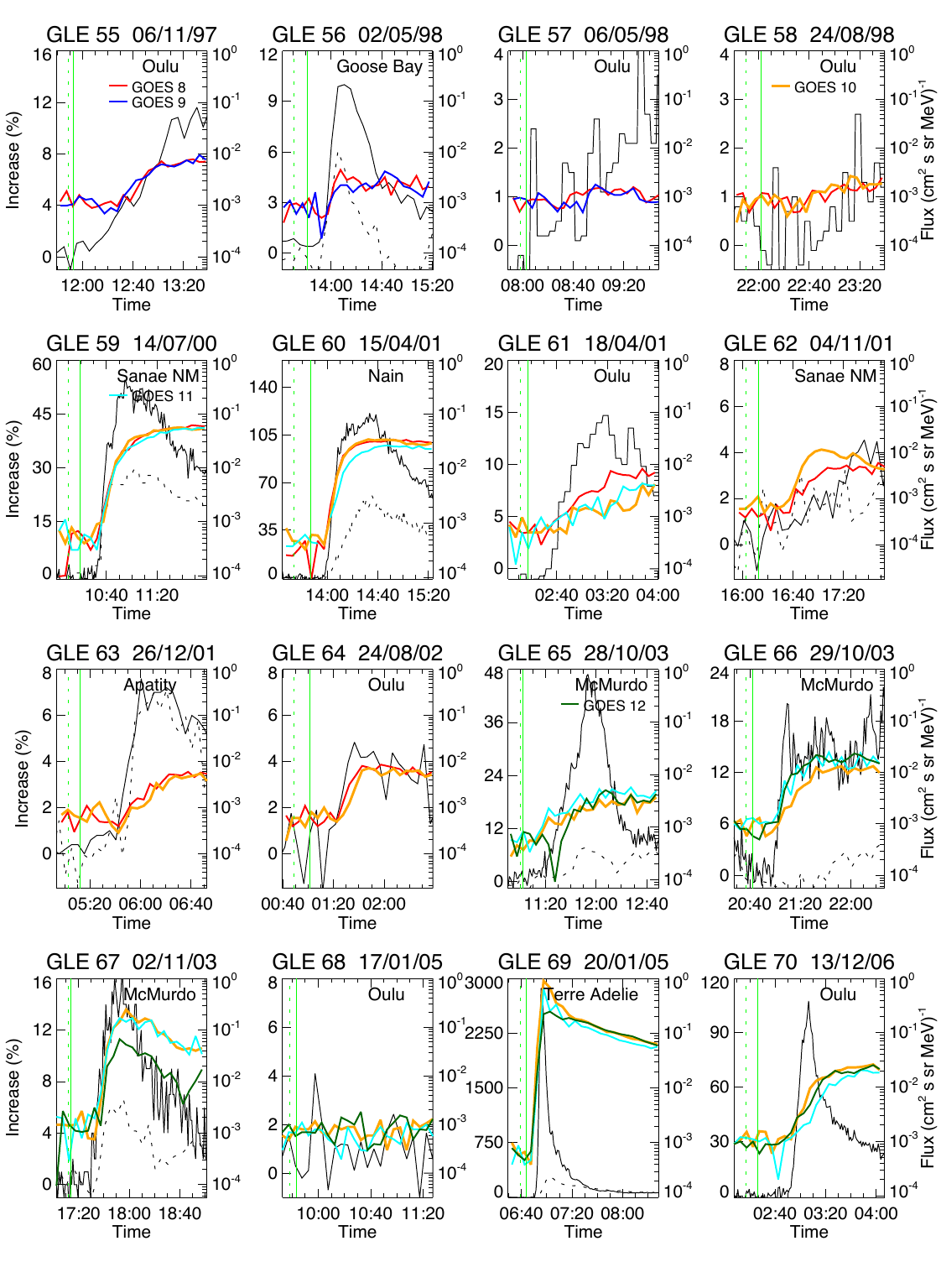}
\caption{Comparison of 16 GLEs in Solar Cycle 23 observed by NMs (in
  black) and GOES HEPAD (in color, P9 channel (420\,--\,510~MeV)).  
  The latter are from GOES 8 and 9 for GLEs 55\,--\,57,
  GOES 8 and 10 for GLEs 58 and 62\,--\,64, 
  GOES 8, 10 and 11 for GLEs 59\,--\,61, and GOES 10\,--\,12 for 
  GLES 65\,--\,70. The primary satellite was GOES~9 for
  GLEs~55\,--\,57, GOES~8 for GLEs~58\,--\,64, and GOES~11 for 
   GLEs~65\,--\,70.
 The name of the NM is shown.  If it is not Oulu, 
  Oulu data are plotted additionally in dotted lines. Green lines indicate the
  onsets of the associated flare (dotted line) and the earliest
  reported type II burst
  (solid line). }
\label{fig:1}       
\end{figure*}

In most GLEs we confirm that the NM and HEPAD onset times
agree to 10 minutes.  
Furthermore, the onset times from the Oulu NM are generally
within five minutes of those that are earliest.  GLEs 65 and 66 are
the only exceptions.
In contrast, the peak fluxes are often
significantly smaller at Oulu than at those NMs that record the
earliest onsets (e.g., GLEs 59, 60, and 69).  

There are two weak GLEs that may be problematic.  While the GLE proton
flux usually exceeds 2$\times$10$^{-3}$ (cm$^{2}$ s sr MeV)$^{-1}$ in the
HEPAD P9 channel, GLE 57 appears to have a smaller increase.  In GLE 68,
it is difficult to find an increase in NM data.  The short peak around 10 UT
may not be due to particles from the Sun, because both the X-ray flare and the
lower-energy proton flux show a slow rise to the peak.  Note that in
a recent study of SEP events in Solar Cycle~23, \cite{Cane10}
indicated that in these events the proton spectra 
did not extend to the GLE ranges.

\begin{figure*}
  \includegraphics[width=1.0\textwidth]{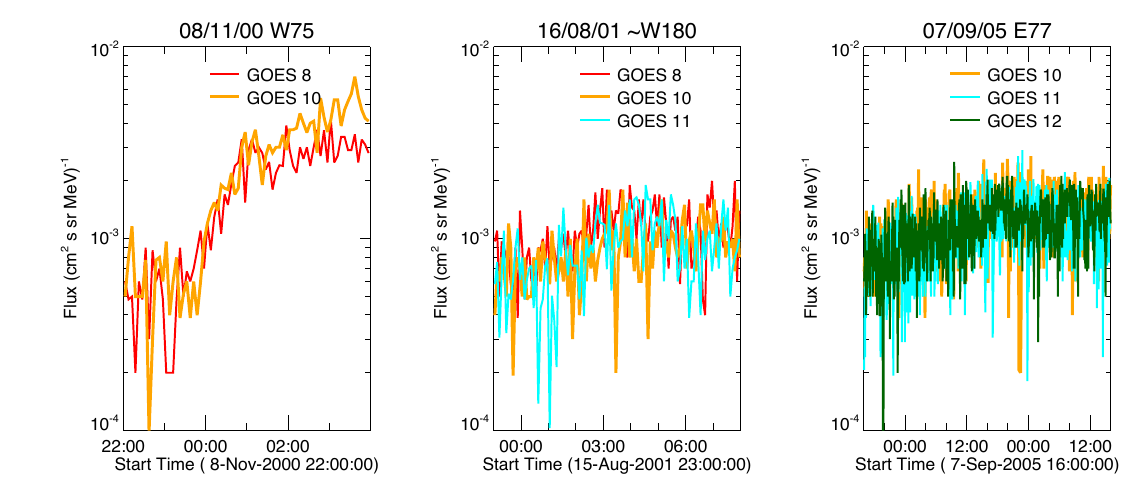}
\caption{Examples of non-GLE events with enhanced 
420\,--\,510 MeV proton flux as
  observed by GOES HEPAD.  }
\label{fig:2}       
\end{figure*}

Although it is possible that the energy of protons has to be higher 
than 500 MeV to produce secondary particles, 
we point out three non-GLE SEP events that have 
the proton flux in P9 channel 
higher than or at least comparable to that of GLE 57.
They are shown in Figure~2. 
The 8 November 2000 SEP event 
represents the third largest $>$30~MeV proton
intensity during Solar Cycle 23 \citep{Mewaldt11}, and it was produced
by a CME from an area between minor ARs \citep{Nitta03}.  
GOES data at 165\,--\,500~MeV and 420\,--\,510~MeV suggest that this
event has a much softer spectrum above $\sim$200~MeV than a majority
of GLEs.
In the two additional events, the proton flux is much smaller and 
increases on longer time scales.  
The source longitude of the 16 August 2001 event
is believed to be near W180 \citep{Cliver05_icrc}, but the
$>$10~MeV proton flux was almost 500 pfu.  The AR responsible for the 7
September 2005 event was close to the east limb, and the $>$10~MeV
protons were seen to increase to $\sim$100~pfu during the first 24
hours.  It is not clear how NMs could observe small events with
gradual time profiles, since the relation of the asymptotic look angle
of a given NM with the interplanetary magnetic field is expected to change 
at least on a time scale of hours.  

These cases indicate
that the distinction of GLEs and non-GLE SEP events may not be clear.
Nevertheless, we shall treat the 16 official GLEs during Solar Cycle
23 as special events.  While GLEs with high intensity
permit us to study their detailed time structures \citep[e.g.][]{Moraal11},
it is difficult even to find the onset time 
if the increase is much less than $\sim$10\%
as in GLEs 57, 58 and 68.  For completeness, however, we keep all the
GLEs in Table 1 for discussion in the following sections.

\section{Properties of the flares and CMEs associated with GLEs}
\label{sec:3}

Apart from a small number of unusual cases \citep{Cliver06}, 
GLEs tend to be associated with intense flares and
energetic (fast and wide) CMEs \citep{Gopal10,Gopal11a}.  Here we
show that they do not serve as sufficient conditions for GLEs. 
Columns 5\,--\,6 of Table~1 show the peak flux and location 
of the flares associated with GLEs.  It is a common practice to 
characterize a flare in terms of its soft X-ray
(1\,--\,8~\AA) peak flux as
measured by the {\it GOES} X-ray Spectrometer (XRS). 
Except for GLE 61, 
whose source region is estimated to be $\sim$26$\arcdeg$ behind the west limb
\cite[see][]{Hudson01}, all the remaining 15 GLEs are associated with flares
above the M7.1 level (7.1$\times10^{-5}$~W/m$^{2}$), and 11 of them are above
the X2 level (2.0$\times10^{-4}$~W/m$^{2}$).  We first compare flares
associated with GLEs with intense flares in general.  

Figure~3(a) is a scatter plot between the X-ray fluxes at
the peak and integrated between start and end times (as taken from the
NOAA event list), which may be
referred to as fluence.  The plotted data consist of all the 15 GLEs from
disk regions and all the 29 flares during Solar Cycle 23 whose peak flux
exceeded the X2 level and which came from the western hemisphere
and slightly behind the west limb. 
The data points for GLEs are plotted in four different sizes on the
basis of the Oulu NM peak increase; 1. $<$6\%, 2. 6\,--\,20\%,
3. 20\,--\,100\%, and 4. $>$100\%.  The count of one NM does not
translate to a physical quantity such as the proton flux at a fixed
energy range (see Section 2).  But this
designation, as used also in Figures~3(b), 4 and 7, may give a 
general idea of the relative magnitudes of our
GLEs.  The symbol of ``possible SEPs'' indicates that 
an SEP event is not clearly isolated due to 
the residual flux from the previous event.  Note
that GLEs 58 and 65, plotted in triangles, were from the eastern hemisphere.  

\begin{figure*}[h!]
  \includegraphics[width=1.0\textwidth]{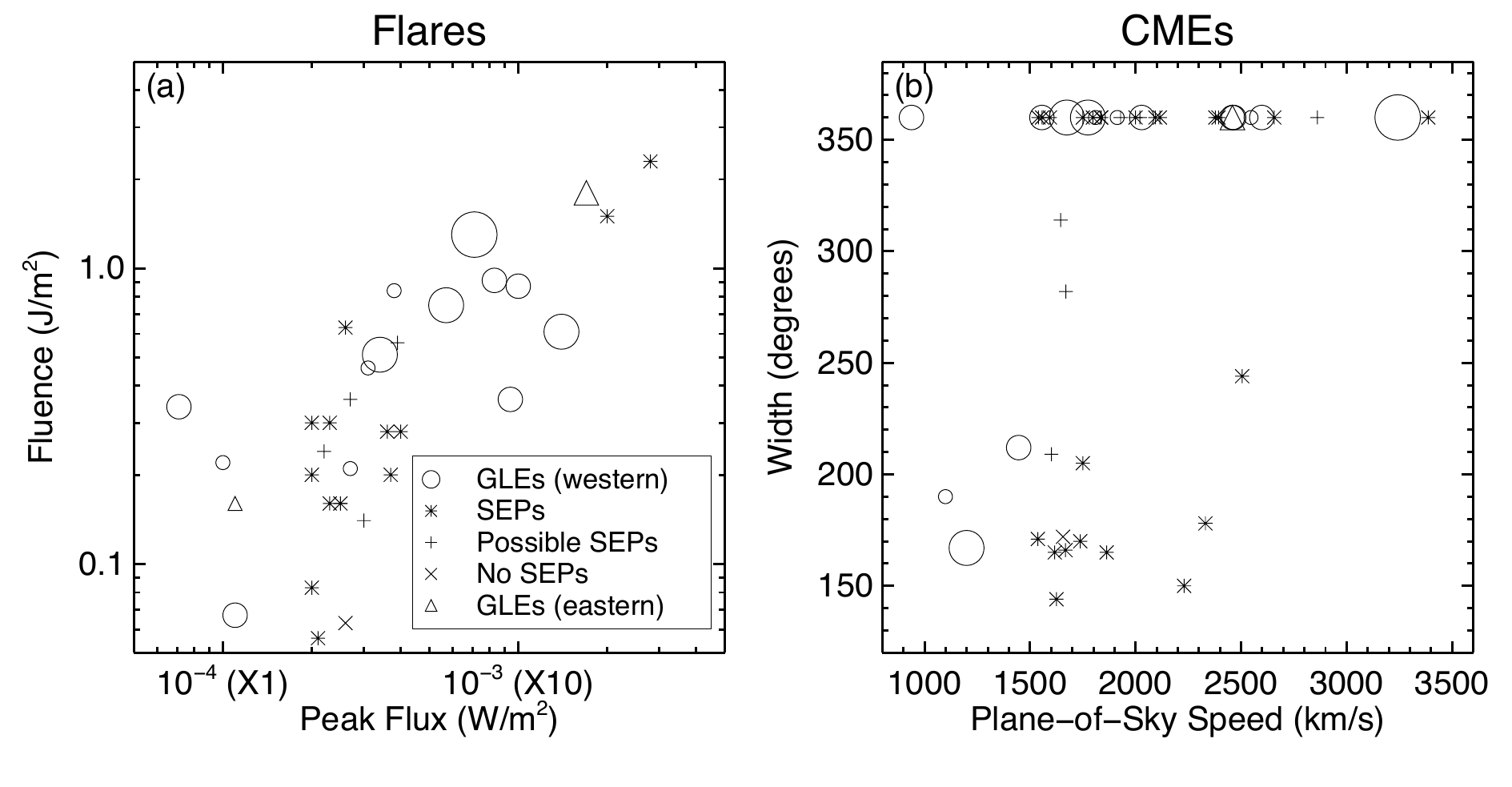}
\caption{Scatter plots of two parameters of flares and CMEs that may 
be correlated with gradual SEPs.  Panel (a) include all the 29 flares 
during Solar Cycle 23 that occurred in 
the western hemisphere and had the peak flux exceeding the X2 level 
(2$\times10^{-4}$~W/m$^{2}$), three less intense flares associated with
GLEs and two flares from the eastern hemisphere.  
Panel (b) shows 41 CMEs faster than 1500~km/s and wider than 140$\arcdeg$ 
from source regions in the longitude range of W00\,--\,W120, 
supplemented by four slower CMEs associated with GLEs 
and the 28 October 2003 event (the only triangle).  
The data points for GLEs are plotted in four sizes on the basis of 
the intensities measured by the Oulu NM (see text). }
\label{fig:3}       
\end{figure*}

There is a good overlap between intense flares and SEP events.  For example, 
if the flare
is from the western hemisphere and above the X2 level, it is almost
always associated at least with a SEP event.  But
only 10 of the 29 such flares are associated with GLEs.  The two
most intense flares are not GLEs; they are the X28 flare on 4 November
2003 at 19:29~UT from AR 10486 located at S19W83,
and the X20 flare on 2 April 2001 at 21:32~UT from AR 09393 located at
N12W82.  Furthermore, four of 15 GLEs are below the X2 level, and there are
a number of flares between X1 and X2 (not plotted) that have no SEPs.
Although it has been known for long time that 
flares that last long, namely long duration
events (LDEs), are more intimately
associated with CMEs \citep{Sheeley75,Kahler77}, 
it appears that the fluence is not a much better indicator of GLEs than the
peak flux.
Part of the reason that the fluence does not make much difference 
may be that the definition of the end time 
as adopted by NOAA, i.e.,
one half of the peak flux during decay, makes it difficult to distinguish
between impulsive flares and LDEs.

Next we discuss whether fast and wide CMEs can serve as sufficient
conditions for GLEs.
The two rightmost columns of Table~1 give the speed and width of the
CMEs as observed by the Large Angle Spectroscopic Coronagraph (LASCO)
\citep{Brueckner95} on the Solar and Heliospheric
Observatory ({\it SOHO}).  Except for the speed of the CME associated
with GLE 69, the values are taken from the CDAW CME Catalog 
(\url{http://cdaw.gsfc.nasa.gov/CME\_list}).
All the CMEs associated
with GLEs are (at least partial) halo CMEs with 
the smallest width of 167$\arcdeg$, 
but the directly  measured speeds, projected in the plane of the sky, 
have a wide range, so do the projection-corrected speeds
\citep{Gopal10,Gopal11a}.  

In Figure~3(b) we compare the CMEs associated and not associated with
GLEs in a plot of the width against the plane-of-the-sky speed.
In addition to all the 15 CMEs associated with GLEs, we include all
the 41 CMEs in Solar Cycle 23 that are wider than 140$\arcdeg$ and faster
than 1500~km/s, with the unambiguously identified source regions in the
western hemisphere or less than 30$\arcdeg$ behind the west limb.  A
total of ten CMEs are overlapped in the two groups.
Four CMEs associated with GLEs are slower than
1500~km/s.  They are not special in terms of the locations, i.e., their
source regions are not necessarily near disk center, or in terms of
the presence of a prior CME, which may help less energetic eruptions
be associated with large SEP events \citep{Gopal04, Cliver06}.  
We therefore conclude that the fast and wide CMEs do not
provide sufficient conditions for GLEs, even though many of them are
associated with gradual SEP events observed at lower energies, 
such as the two most intense flares mentioned earlier that are 
associated with halo CMEs with speeds of 2657~km/s and 2505~km/s.

\begin{table}
\caption{Basic properties of the 11 ARs associated with GLEs} 
\begin{tabular}{rrrrrrrr}
\hline\noalign{\smallskip} 
\multicolumn{1}{c}{} & \multicolumn{1}{c}{Date of} &
\multicolumn{1}{c}{} & \multicolumn{1}{c}{Max} &
\multicolumn{1}{c}{Mag} & \multicolumn{1}{c}{Sunspot} &
\multicolumn{1}{c}{USF} & \multicolumn{1}{c}{Longitude and} \\
\multicolumn{1}{c}{AR} & \multicolumn{1}{c}{CM passage} & \multicolumn{1}{c}{Age$^{a}$} & \multicolumn{1}{c}{Area$^{b}$} & \multicolumn{1}{c}{Type} & \multicolumn{1}{c}{Rotation} & \multicolumn{1}{c}{PIL$^{c}$} & \multicolumn{1}{c}{(Area$^{b}$)} at GLE \\
\noalign{\smallskip}\hline\noalign{\smallskip}
08100 & 1997/11/02 & $<$1 & 1000 &  $\beta$-$\gamma$-$\delta$ & N & 0.35 & W63 (840)     \\
08210 & 1998/05/01 & $<$1 & 480 & $\beta$-$\gamma$-$\delta$ & Y & 0.20 & W15 (390)  \\
      &            &     &     &                           &   &      & W66 (450)  \\
08307 & 1998/08/26 & $<$1 & 570 &  $\beta$-$\delta$ & N & ?$^{d}$ & E09 (380)   \\
09077 & 2000/07/14 & $<$1 & 1010 &  $\beta$-$\gamma$-$\delta$ & Y & 0.63 & W07 (620)   \\
09415 & 2001/04/09 & $<1$ & 790 &  $\beta$-$\gamma$-$\delta$ & Y & 0.45 & W85 (350) \\
      &            &     &     &                             &   &      & W116 (?)  \\
09684 & 2001/11/03 & $<1$ & 550 &  $\beta$-$\gamma$-$\delta$ & Y & 0.18 & W18 (510)  \\
09742 & 2001/12/22 & $<1$ & 1070 &  $\beta$-$\gamma$-$\delta$ & Y & 0.29 & W54 (900)   \\
10069 & 2002/08/18 & 2-3 & 1960 &  $\beta$-$\gamma$-$\delta$ & Y & 0.70 & W81 (800)  \\
10486 & 2003/10/29 & $<1$ & 2610 &  $\beta$-$\gamma$-$\delta$ & Y & 0.77 & E15 (2150) \\
      &            &     &     &                           &   &      & W02 (2250)  \\
      &            &     &     &                           &   &      & W59 (2080)  \\
10720 & 2005/01/16 & $<1$ & 1630 &  $\beta$-$\delta$ & Y & 1.04 & W25 (1570)  \\
      &            &     &     &                           &   &      & W61 (1290)  \\
10930 & 2006/12/12 & $<1$ & 680 & $\beta$-$\gamma$-$\delta$ & Y &  0.32 & W23 (680) 	\\

\noalign{\smallskip}\hline
\noalign{\bigskip}

\end{tabular}

\noindent
a. In solar rotation. 

\noindent
b. In microhemispheres.  Corrected for foreshortening. 

\noindent
c. At maximum, in the unit of 10$^{22}$ maxwell.

\noindent
d. No SOHO MDI data.
\end{table}

\section{Basic properties of the ARs associated with GLEs}
\label{sec:3}

This section considers the possibility that the ARs associated with
GLEs may have
unique properties.  There are 11 ARs that produced the CMEs associated with
GLEs during Solar Cycle 23.  Table~2 gives their basic properties, i.e.,
age, sunspot area, magnetic type, detection of sunspot rotation, 
unsigned magnetic flux near the polarity inversion lines.  
The age of an AR can be determined quite easily to the time
scale of a solar rotation by locating it on synoptic magnetograms in successive 
Carrington rotations.
Most (10/11) of the GLE ARs are less than one solar rotation
old, even though some of them survived a few more rotations after the GLEs,
such as AR~08100 \citep[see][]{Green02}.  Six of the GLE ARs were seen
to emerge on the visible disk, but it was at least a week before a GLE occurred.
According to \cite{Harvey93}, the lifetime tends to be longer
for larger ARs, and the ARs associated with GLEs tend to be large.
Therefore, an old age could have been a discriminating factor for GLE
ARs, but most of the GLEs in Solar Cycle 23 
occurred in early stages of AR evolution.

Next, we study the sunspot area from the NOAA daily AR list also
available in SolarSoft.  The
fourth column of Table~2 shows the maximum sunspot area during the
region's disk passage.  
One of the GLE regions, AR~10486, which was also the progenitor of
severe space weather during the Halloween (October\,--\,November) 2003
period, had
the largest sunspot area of all ARs in Solar Cycle 23, 
but five GLE ARs are smaller than 800
microhemispheres at their maxima.  They are even smaller around the
times of the GLEs, as shown in the last column of Table~2, which are
interpolated from the two measurements closest to the GLE times.  Therefore,
the large sunspot area does not serve as a discriminating factor for
GLEs. According to \citet{Gopal05_iau}, the sunspot area is less
strongly correlated with the SEP intensity and CME 
speed than with the flare intensity.   

The parameters in columns 5\,--\,7 are more susceptible to 
foreshortening as the AR
approaches the west limb.  Therefore the information comes from the
time when the AR's longitude was less than 60$\arcdeg$.  
Column 5 shows the magnetic type of the ARs, following the Mt. Wilson
classification scheme \citep{Hale19, Kunzel60}.
It 
refers to the most complex type the AR undergoes during its
evolution.  As expected, all the GLE regions have magnetic type that
contains $\delta$ spots at least during some evolutionary stages.
We have also studied sunspot rotation \citep{Brown03} for the GLE
regions, using white-light images from the Transition Region and
Coronal Explorer (TRACE) \citep{Handy99} or magnetograms from the 
Michelson-Doppler Imager (MDI) \citep{Scherrer95} on {\it SOHO.}
Although this dynamic phenomenon is frequently associated with
major flares \citep[e.g.][]{Nightingale05,Zhang08,Kazachenko09}, 
it is missing or at least unclear
in two of the 11 GLE ARs.  Moreover, there are many ARs that
show sunspot rotation in the western hemisphere but do not accompany 
gradual SEP events, not to mention GLEs.

Another indicator for major flaring comes from measurement of
photospheric magnetic field.  One of them is
the total unsigned flux (USF) close to the polarity inversion lines (PILs).  
We identify regions with strong field of opposite polarities
tightly packed by multiplying magnetograms that are dilated with a kernel of
3$\times$3 MDI 2$\arcsec$ pixels \citep[see][]{Schrijver07}.
The flux in maxwell (Mx) is shown rather than Schrijver's parameter 
{\it R}, which is essentially the flux divided by the area corresponding
to the MDI pixel (2.2 $\times$10$^{16}$~cm$^{2}$).  All the 10 GLE ARs
observed by MDI have the USF near the PILs
above a threshold of 2$ \times 10^{20}$Mx, which accounts for 99\% of 
X-class flares (see Figure~3 of Schrijver 2007).  
But many other ARs belong to this
category, as shown below, so this parameter may not be used to predict GLEs.

\begin{table}
\caption{Comparison of AR parameters (average and standard deviation) 
in three categories} 
\begin{tabular}{lrrr}
\hline\noalign{\smallskip} 
\multicolumn{1}{c}{} & \multicolumn{1}{c}{GLE regions} &
\multicolumn{1}{c}{SEP regions} & \multicolumn{1}{c}{$\delta$ regions
  (no SEP)} \\
\noalign{\smallskip}\hline\noalign{\smallskip}
Sunspot area (microhemispheres) & 1102$\pm$621 & 633$\pm$476 & 646$\pm$303 \\
Total USF (AR) (10$^{22}$ Mx) & 11.9 $\pm$5.1 & 9.0$\pm$4.4 & 9.5$\pm$3.5 \\
Total USF (PIL) (10$^{21}$ Mx) & 4.9 $\pm$2.8 & 3.5$\pm$2.4 & 2.7$\pm$1.3 \\
Net Flux (AR) (10$^{22}$ Mx) & 1.5 $\pm$0.7 & 1.7$\pm$1.6 & 1.2$\pm$1.0 \\
Separation (10$^{9}$ cm) & 3.8 $\pm$2.1 & 4.7$\pm$2.6 & 7.0$\pm$2.7 \\

\noalign{\smallskip}\hline
\noalign{\bigskip}

\end{tabular}
\end{table}

\begin{figure*}
  \includegraphics[width=1.00\textwidth]{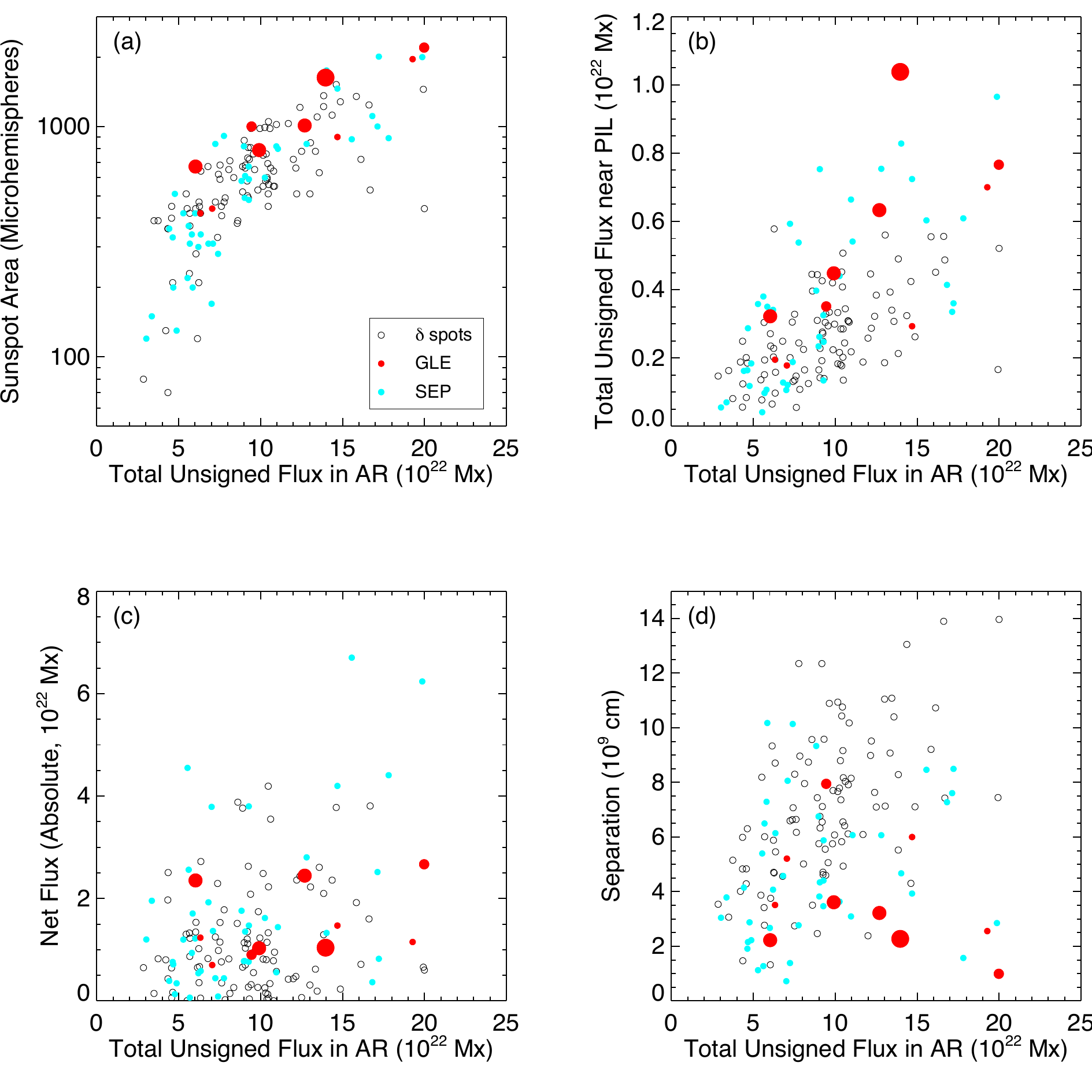}
\caption{Four quantities plotted against the total 
USF of
  the whole AR, separately for ARs associated with GLEs
and non-GLE (lower-energy) SEPs, and $\delta$ regions not associated
with either of them.  They are (a) sunspot area, (b) total
USF near PILs, (c) net flux (in absolute value) of the
whole region, and (d) separation of the centroids of positive and
negative polarity regions.  As in Figure~3, 
the data points for GLEs are plotted in four sizes on the basis of 
the intensities measured by the Oulu NM (see Section~3). }
\label{fig:4}       
\end{figure*}

Now we compare the properties of ARs that produced GLEs
with those of other ARs in a large sample.  We have
systematically studied the properties of ARs 
in the mission-long MDI full-disk magnetograms
which are sampled every 96 minutes, in a similar manner to the recent study by
\cite{Mason10}.  In addition to the ten GLE ARs,  
we extract a total of
34 regions that produced SEP events (but not GLEs) and 96 non-SEP
$\delta$ regions in the western hemisphere.  Table~3 gives average
values 
of the parameters we are concerned with.  They are: the sunspot
area, the total USF in the whole region and near the
PILs, the net magnetic flux of the whole AR, and 
the separation of the centroids of the two polarity areas.  
The parameters for a given
region come from the magnetogram in which the unsigned total flux of
the whole region is largest while the region is within the $\approx$
60$\arcdeg$ longitudinal range.  Note that this does not correspond
to the GLE time.  The sunspot area is again taken from the NOAA daily active
region list for the day closest to the magnetogram time.

Figure~4 gives scatter plots of four of the AR parameters in Table~3
(sunspot area, etc.) vs the fifth (total USF).
As in Figure~3,
the data points for GLEs are plotted in four different sizes reflecting 
the peak intensities in the Oulu NM data.  From Table~3 and
Figure~4, the following trends, although not statistically
significant, can be found.  
The GLE ARs have larger
sunspot area and total USF both in the whole AR and near the PILs.
In contrast, the high net flux may be more characteristic to the SEP
regions than to the GLE regions.  Finally, the GLE regions tend to be
more compact than $\delta$ regions without SEPs, 
even though regions with larger separation tend to
produce more intense flares and energetic CMEs \citep[e.g.,][]{Guo06}.  
Despite the above trends, these basic properties may not be used to
distinguish GLEs due to large uncertainties, and more
importantly to the time difference between the measurements and the GLEs.
In short, none of the basic properties of ARs may be considered to be
discriminating factors for GLEs.  
There is still a possibility that
more advanced
observables of ARs can discriminate GLEs.  For example, pronounced
changes of horizontal magnetic field were reported for AR~10720 just
before GLE~69 occurred \citep{Wang09}.
However, these advanced observables were available 
only for a handful of ARs 
during Solar Cycle 23.

\section{Magnetic field connection}
\label{sec:4}

In the previous sections, we have shown that neither the magnitude
of flares and CMEs nor the basic properties of ARs can be used to
distinguish GLEs.  Now it is worthwhile to turn to other factors that
may contribute to the detection of GLEs. In this section, we discuss 
the possible importance of magnetic field connection of the source ARs 
and surrounding areas. In the context of SEP events, the western hemisphere,
centered around the longitude of W60, is usually thought of 
as being ``well-connected.''  This is because the interplanetary magnetic field
is approximated as the Parker spiral, and the ends of the field lines 
that come to the Earth are mapped to the solar longitude of W60 
for a typical slow wind speed of $\approx$400~km/s.  
We expect higher particle flux and quicker onset in SEP
events from well-connected regions.
At first, we note that some GLEs in
Solar Cycle 23 were indeed located around the well-connected longitudes  
(GLEs 55, 57, 63, 67, and 69).  In table 4, the second column
replicates the coordinates of the flares associated with the GLEs.  
In particular, the flare associated with GLE 69, which had an unusually 
high flux and quick onset, occurred around W60.

\begin{table}
\caption{GLEs and magnetic field connection} 
\begin{tabular}{lrcrrrrrrrccc}
\hline\noalign{\smallskip} 
(1) & (2) & (3) & (4) & (5) & (6) & (7) & (8) & (9) & (10) & (11) \\
 & Flare & & PS  & Diff. & PFSS & Dist. & & & Delay & \\
 & Coord. &  $v_{SW}$ & Long. & Long. & Coord. & $\Delta d$ & Pol. &
Pol. &  $\Delta t$ & ICME$^{b}$ \\ 
ID & ($\phi$, $\lambda$)$_{f}$ & (km/s) & $\phi _{PS}$ & $\Delta \phi$ & 
($\phi$, $\lambda$)$_{PF}$ & $|$f-PF$|$ & Sun & SW & (minutes) &
(hours) \\  
\noalign{\smallskip}\hline\noalign{\smallskip}
55       & 63, -18 & 350 & 65 & -2  & 63, -18 & 0 & M & pM & 21 & -11  \\
56       & 15, -15 & 601 & 38 &-23  & 14, -17 & 2 & M & M &  24 & I  \\
57       & 66, -11 & 466 & 50 & 16  & 64, -16 & 6 & M & M &  27 & I  \\
58       & -9,  35 & 423 & 55 &-64  & 25, 39$^{a}$ & 28 & P  & P & 60 & N \\
59       &  7,  22 & 604 & 39 &-32  & 26, 13 & 20 & P & Pm & 20 & I \\
60       & 85, -20 & 517 & 45 & 41  & 63, -5 & 26 & M & pM & 26 & -3$^{c}$  \\
61       & 116, -20& 483 & 48 & 68  & 57, -10 & 58 & M & pM & 24 & I$^{d}$  \\
62       & 18, 6   & 304 & 74 & -57 & 61, 10 & 43 & M & Pm & 57 & N  \\
63       & 54, 5   & 381 & 59 & -5  & 60, 12 & 10 & P & pM & 27 & N  \\
64       & 81, 2   & 387 & 60 & 21  & 85, -8 & 7  & M & M &  29 & N  \\
65       & -8, -18 & 772 & 29 &-37  & 54,  3 & 64 & M & M &  22 & +2  \\
66       &  4, -18 &1125 & 20 &-16  & 14,-14 & 11 & M & P &  53 &  I  \\
67       & 57, -18 & 515 & 44 & 13  & 53,-19 &  4 & M & pM & 18 &  N  \\
68       & 25,  14 & 678 & 33 & -8  & 21, 15 &  4 & M & P &  17 & +3$^{e}$ \\
69       & 61,  14 & 848 & 27 & 34  & 55, 12 &  6 & M & pM & 12 & +4  \\
70       & 23,  -6 & 648 & 35 &-12  & 29, -4 &  6 & M & M &  28 & N  \\

\noalign{\smallskip}\hline
\noalign{\bigskip}
\end{tabular}

\noindent
a. Use Kitt Peak data, because no MDI data are available.

\noindent
b. Refer to \cite{Richardson10}.  ``I'' and ``N'' stand for ``inside an
ICME'' and ``no ICME'', respectively.  The positive (negative)
numbers refer to hours after the end (before the start) of the ICME.

\noindent
c. Minor ICME.  CME not identified, which is likely from a different region.

\noindent
d. CME not identified, which is likely from a different region.

\noindent
e. CME from a different region (AR 10718)

\end{table}

However, as we take into account the observed solar wind speed around the GLE
time as shown in the third column of Table 4, the footpoints of the
well-connected field lines are sometimes at widely different
longitudes (column 4).  For example, the difference in longitude 
between the flare and the footpoint of the well-connected 
Parker spiral field line (column 5) is now as much as 
34$\arcdeg$ for GLE 69.  
This would seem to imply that the detection of GLEs may not
depend on the magnetic field connection, as long as we assume that the
Parker spiral goes down to the photosphere.  However, this picture  
should not be realistic.
We never see coronal images that are dominated by simple radial
magnetic field which, in combination with solar rotation, results in the 
Parker spiral.  Images in soft X-rays and EUV reveal the solar corona
highly structured with closed loops which likely trace magnetic field
lines, and with dark regions, some of which correspond to coronal holes or
open field regions.  Moreover, the concept of 
the ``well-connected longitudes'' ignores 
the latitudinal excursion of field lines that is needed for a
non-ecliptic AR to be connected to the Earth.

Here we take another step and conduct a simple modeling of coronal
magnetic field, using the potential field source surface
(PFSS) model \citep{Altschuler69, Schatten69}.  This model assumes
that there is no current in the corona below the source surface,
usually a sphere of radius 2.5~$R_{\sun}$, on which the magnetic
field is made radial into interplanetary space.  
That is, open field lines are those that 
reach the source surface.  Well-connected field lines are a subset of
open field lines that intersect the source surface at the ecliptic.
We adopt the PFSS model, as implemented into 
SolarSoft by \citet{Schrijver03}, which is based on the 
photospheric magnetic maps that are constructed from 
(1) MDI data for the area within 60$\arcdeg$ from
disk center and (2) the flux-dispersal model \citep{Schrijver01} for
the remainder of the solar surface.  Such evolving magnetic maps are
generated and PFSS extrapolations calculated every 6 hours.

Following \citet{Nitta06}, we trace field lines first from Earth
to the source surface assuming the Parker spiral, and then to the
photosphere using the PFSS model.  We allow for $\pm$7.5$\arcdeg$ and 
$\pm$2.5$\arcdeg$ uncertainties for the longitude and latitude,
respectively, of the source surface coordinates, in an attempt to address
possible deviations of the interplanetary magnetic field from 
the Parker spiral.  We identify the field line whose
footpoint is closest to the flare.  The coordinates 
 ($\phi$, $\lambda$)$_{PF}$ of the footpoint
of this field line in the photosphere  
and its distance ($\Delta d$) from the flare along a great circle 
are given in the
6th and 7th columns, respectively.  As in the case of
impulsive SEP events (see Figure~8 in Nitta et al. (2006)), there are
more GLE events in which the footpoint of the field line that connects 
to Earth is closer to the associated flare than if we assume 
the Parker spiral all the way to the photosphere
(5th column).

\begin{figure*}[h!]
  \includegraphics[width=1.00\textwidth]{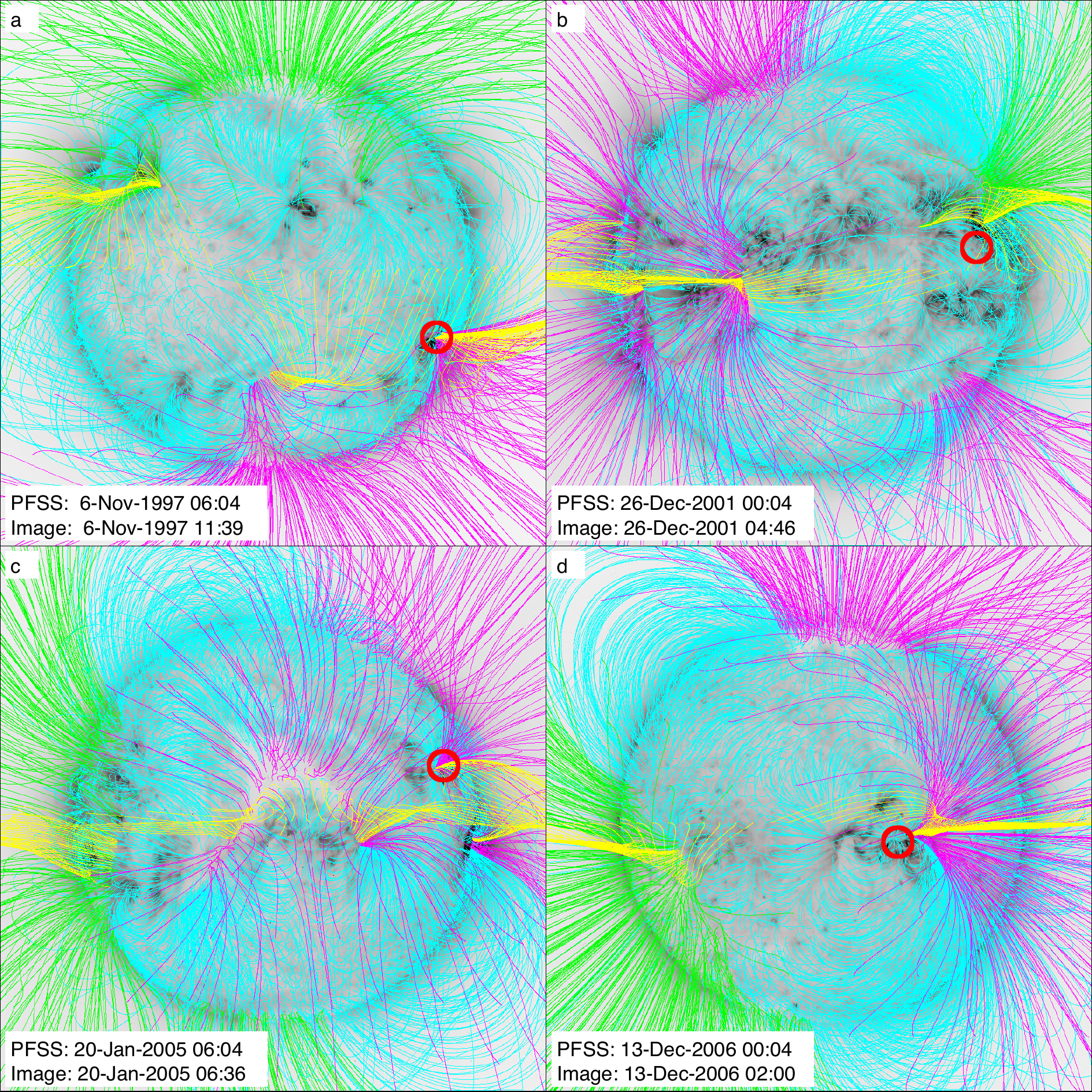}
\caption{Extrapolated magnetic field lines plotted on EIT images in
  reverse color that are taken shortly before four GLEs (55, 63, 69
  and 70).  The PFSS
  model is used for extrapolation.  Lines in cyan are closed field
  lines, whereas those in green and pink are open field lines with
  positive and negative footpoints, respectively.  Those in yellow are
  open field lines that are ecliptic at the source surface, a subset
  of which is expected to be connected to Earth.  The locations
  of the GLE-associated flares are indicated by circles in red.}
\label{fig:5}       
\end{figure*}

In one half of the GLEs, there is a field line whose
photospheric footpoint is within 7$\arcdeg$ from the flare, and whose
coordinates on the source surface are consistent with the observed
solar wind speed.  
We may consider these GLEs to be
well-connected, including GLE 69.  This result suggests that magnetic
field connection of the AR may be an important factor for at least
certain GLEs.  Field lines are traced also upward from the photosphere. 
Figure~5 shows PFSS extrapolations for four well-connected events 
on images from the Extreme-ultraviolet Imaging Telescope
(EIT) \citep{Boudin95} on SOHO.  
Both the images and the extrapolations refer to times slightly
before the GLEs.

In six events, the footpoints of the well-connected field lines are
separated from the flare by more than 
20$\arcdeg$, possibly arguing against the importance of 
magnetic field connection of the AR to Earth for these GLEs.
In two cases (e.g., AR 9415 for GLEs 60 and 61), 
open field lines are found close to the flare, 
as the field is extrapolated from the photosphere, 
but they intersect the
source surface far from the ecliptic. 
In other cases, no open field lines are found around the AR.  
These are labeled as occurring in a closed magnetic field environment.
In
addition to AR 9077 (GLE 59) and 9684 (GLE 62), AR 10486 for GLE 65
belongs to this category.  

\begin{figure*}[h!]
  \includegraphics[width=1.00\textwidth]{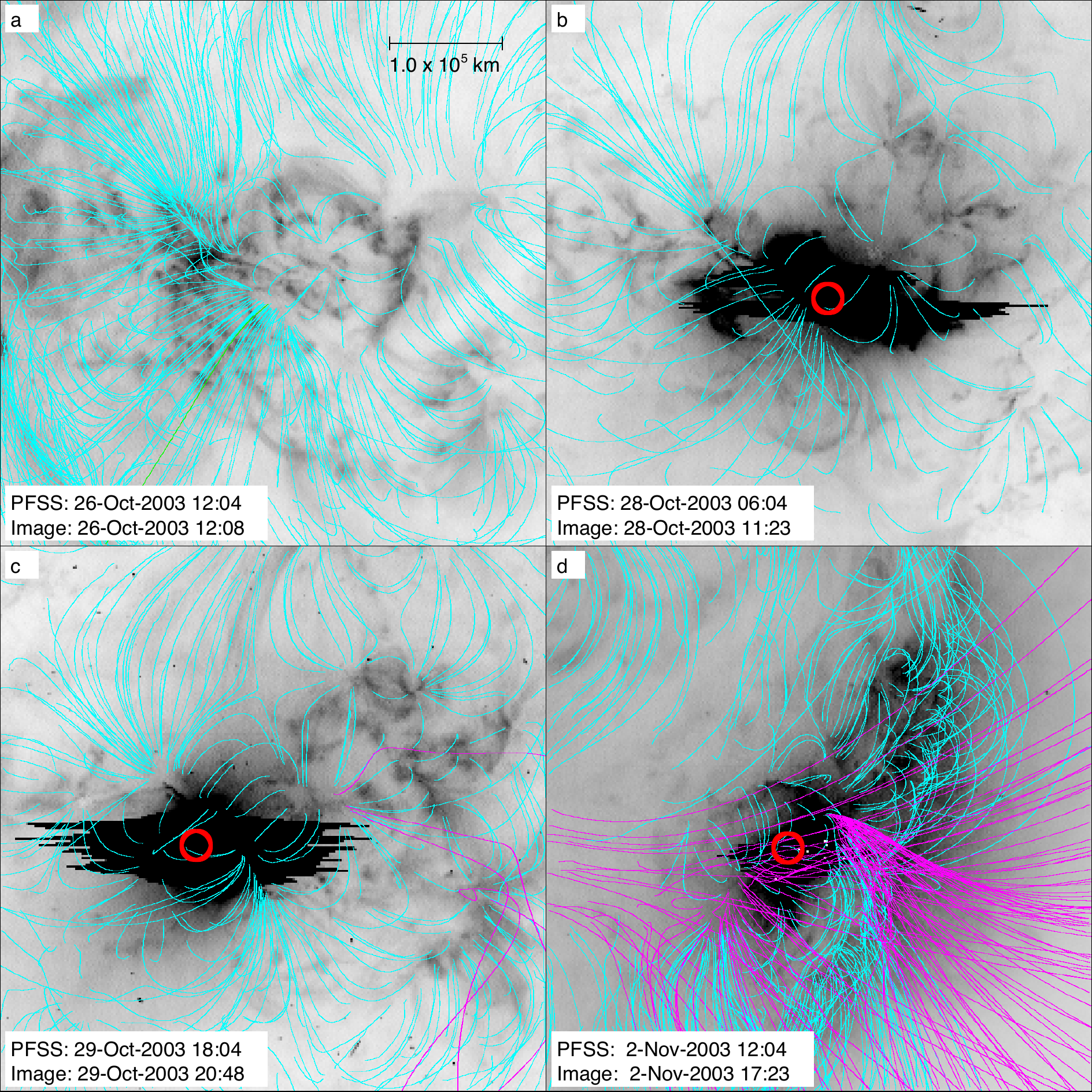}
\caption{Magnetic field extrapolation for AR~10486 and the surrounding
  region at four different
  times.  Field lines are plotted on EIT images taken (a) two days before
  GLE 65, (b)\,--\,(d) around the times of GLEs 65\,--\,67.  The same colors
  are used as Figure~5 except that ecliptic field lines are not
  distinguished. }
\label{fig:6}       
\end{figure*}

Figure~6 shows the evolution of large-scale
magnetic field around AR~10486.  The region is found to be closed
until October 29, when we start to see some open field lines of
negative polarity (Figure~6(c)).  By the time GLE 67 occurred on
November 2, the region was largely open.  It is possible that the PFSS
model does not properly capture the likely complex magnetic topology of ARs as
flare- and CME-prolific as AR~10486, considering its simplifying
assumptions of no current in the coronal volume.  Moreover, it is a
static model, not containing information on dynamics.  
However, let us for now assume that the PFSS model adequately locates 
open field lines in areas around the ARs and that 
these changes reflect the evolution of the ARs and surrounding areas.   
Of course we are keenly aware of  
the need to evaluate the PFSS model in individual
cases \citep[see, for example][]{Nitta08}. 
Incidentally, 
\citet{YangLiu07} found AR~10486 embedded in open field
environment, using less frequently updated synoptic magnetic data.


As a consistency check we also compare the photospheric
polarity of the open field line closest to the flare and the polarity
of the interplanetary magnetic field around the GLE onset.  These are
given in the eighth and ninth columns of Table 4, where `P' stands
for plus (positive or `away' polarity) and `M' stands for minus
(negative or `toward' polarity).  On the Sun one polarity is
assigned by the PFSS model, but at 1 AU, the
polarity sometime changes on time scales of ten minutes, not captured
in hourly data; we examine high time resolution data from
both Wind and ACE.  In such cases, the polarity that lasted longer is
shown as upper case.  This variation of polarity in the interplanetary
magnetic field may indicate departure from a simple Parker spiral, 
and add another complexity we cannot address within our simple
approach.  However, we see a general agreement of the polarities at the
Sun and 1 AU.

To see the effect of magnetic field connection of the source region 
on how quickly the GLE starts after the flare onset, we plot in 
Figure~7(a) $\Delta d$ with the time difference $\Delta t$ 
of the GLE and flare onsets
(column 10 of Table~4).
We hardly find a correlation between $\Delta t$ and $\Delta d$.
One reason for this may be that the CME associated with
the GLE occurs while Earth is within an interplanetary CME (ICME) that
results from an earlier CME from the same source region.  Then the
path length of the field lines that constrain the particles could be
longer than the typical Parker spiral (often assumed to be 
$\approx$1.2~AU long.)
We use filled circles for these GLEs in Figure~7(a).  
In the last column of Table~4 we also indicate those GLEs that were 
within several hours before or after the passage of an ICME,
because interplanetary magnetic field may be disturbed in
those time ranges.  It may account for GLE~55 ($\Delta d$=0, $\Delta
t$=21), which appears to come through disturbed interplanetary
conditions (the polarity alternating).  However, the apparent 
lack of correlation in
Figure~7(a) will not change if we plot only those GLEs clear of ICMEs.
Note that the path lengths for electron events that occur inside ICMEs
are not substantially different from those that occur outside ICMEs 
\citep{Kahler11a}.

\begin{figure*}[!h]
  \includegraphics[width=1.0\textwidth]{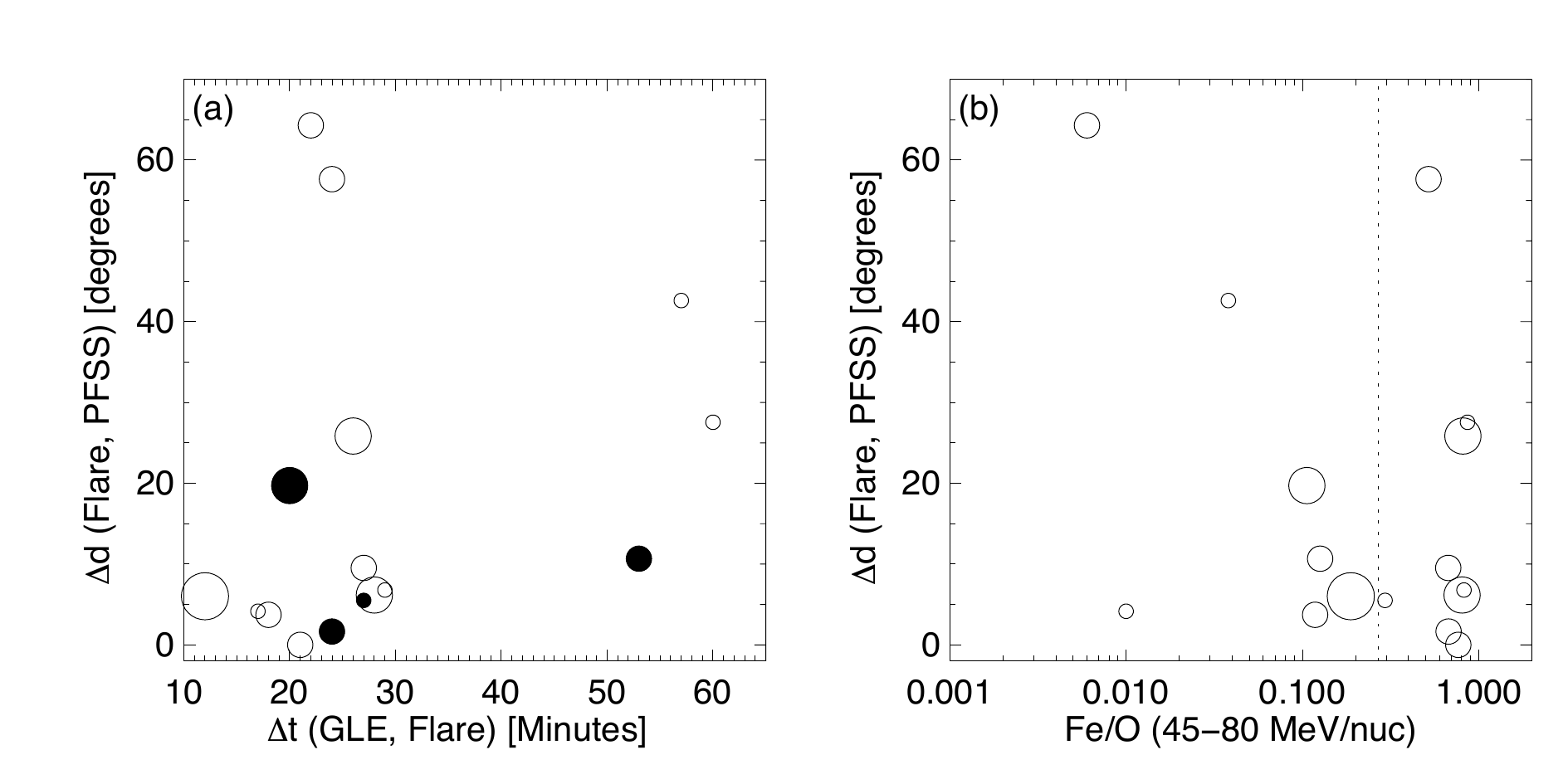}
\caption{The relation of the magnetic field connection with (a)
  the GLE onset time with respect to the flare onset time, and (b) the
Fe/O ratio at 45\,--\,80 MeV/nuc (from \cite{Mewaldt11}). The
right-hand side of the dotted line placed 
at twice the average SEP value (0.134)
of Fe/O \citep{Reames95} may define Fe-rich
SEP events \citep{Tylka05}. In (a),
filled circles are used for GLEs that were inside an ICME. 
The data are plotted in four sizes on the basis of 
the intensities measured by the Oulu NM (see Section~3). }
\label{fig:7}       
\end{figure*}

In Figure~7(b), we plot $\Delta d$ against the Fe/O ratio, whose 
enhancement is sometimes thought to be an indicator of
flare-accelerated particles.  Flares are much less extended than CMEs,
so in order for flare-accelerated particles to be injected on to
well-connected field lines, these field lines need to be rooted 
close to the flare.
Although a few well-connected events are indeed Fe-rich, no
such correlation is seen in the whole sample.  The most intense GLE
(GLE~69), which is well-connected, is not particularly Fe-rich.

\section{Large-scale coronal disturbances}
\label{sec:6}

What can affect the onset times of GLEs?  
They are up to an hour after the flare onsets, 
and these delays are roughly comparable
to those of near-relativistic electron events 
with respect to the associated type III
bursts \citep{Krucker99, Haggerty02}.  \citet{Krucker99}
attributed electron events with delayed onsets, mostly
from what were thought to be poorly-connected regions, to the times for
coronal waves to propagate from the flare region to well-connected
longitudes.  Here, coronal waves were represented by EIT waves 
\citep{Moses97,Thompson98}, which were found to be associated with
electron events with delayed onsets, although the waves
were too slow to account for relatively short delays.

\begin{figure*}
  \includegraphics[width=1.0\textwidth]{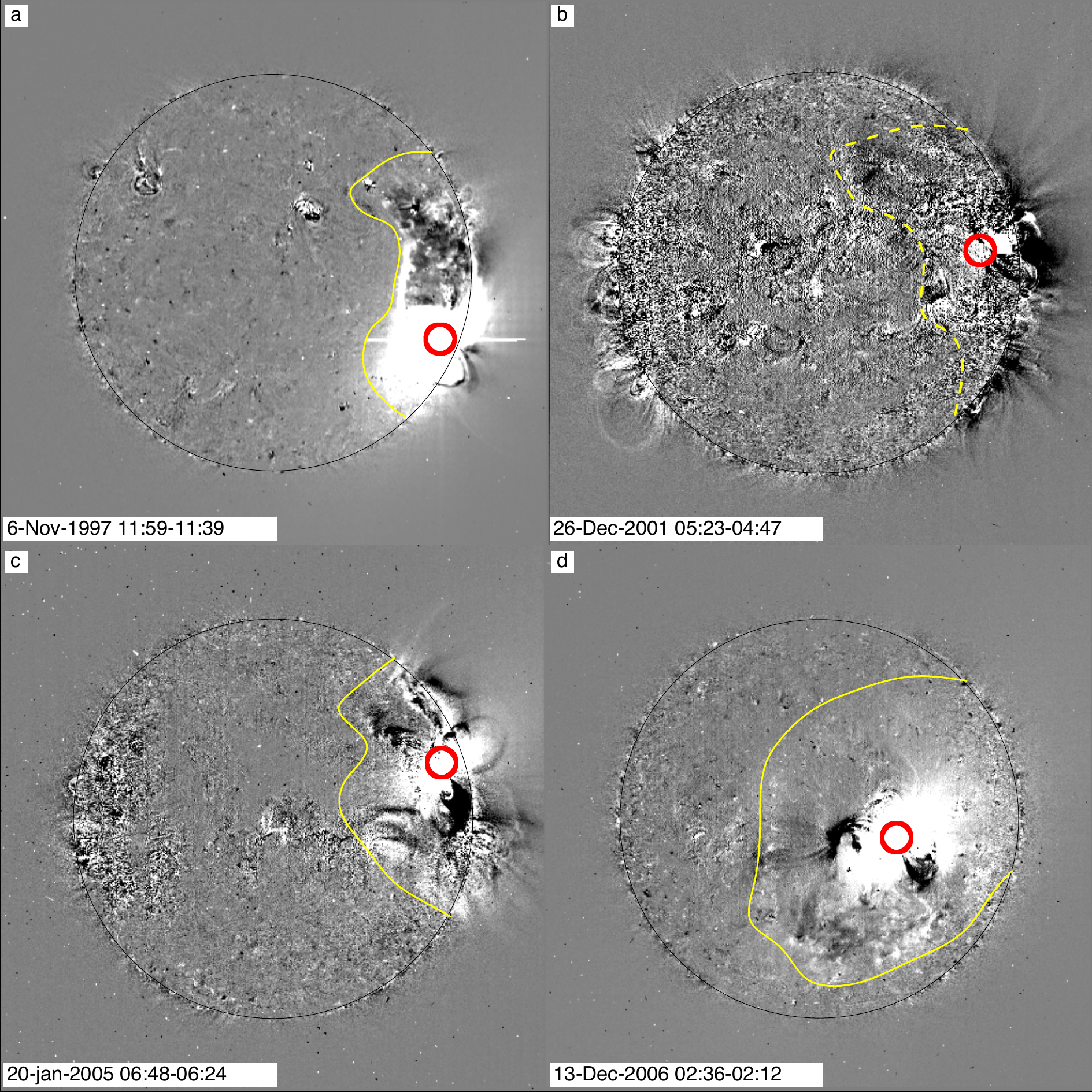}
\caption{EIT 195~\AA\ difference images around the flare
  maximum for the same GLEs as shown in Figure~5 (55, 63, 69 and 70).  
The red circles mark the flare locations.
 Yellow lines outline what appear to be the fronts of the
 disturbances.  In panel (b), a dotted line is used because the front
 is traced only marginally.}
\label{fig:7}       
\end{figure*}

We now look for EIT waves possibly associated with GLEs 
by analyzing EIT images. 
Flares associated with all the GLEs but 58 and 62 were observed by
EIT.  It turns out that we seldom see clear EIT waves around GLE
onsets.  
In Figure~8
we show EIT difference images of the same events as Figure~5.  They
are typically the first image after the flare onset or the one close
to the flare maximum in soft X-rays.  In panels (a)
and (c), distant areas from the flare appear to be already disturbed,
but waves are not found. According to \cite{Thompson09}, who compiled
EIT waves through June 1998, none of the transients associated with
GLEs 55\,--\,57 are identified with EIT waves with high reliability.
If we assume that they start at the flare onset, their speeds are in
the range of 500\,--\,2000~km/s, depending on where they started.  It could be
near the flare center or some 10$^{5}$~km away as in typical Moreton
waves \citep{Moreton60,Warmuth04}.  It would have been difficult 
to observe a wave with a speed of $\gtrsim$1000~km/s
with the typical 12 minute cadence of the EIT.
In the event associated with GLE 63 (panel (b)), the large-scale changes
cannot be detected with confidence.  Therefore a dotted line is used
to indicate the possible front.
GLE 70 seems to be the only GLE associated with a typical EIT
wave whose speed at the bright front is 
400\,--\,450~km/s as measured in two successive
images at 02:36 and 02:48~UT.

In a majority of other GLEs, 
we see large-scale disturbances without clear EIT
wave appearance, similar to panels (a) and (c).  The EIT image that
shows the change in large areas is usually within the time range of
the metric type II radio burst as included in the list
available at the National Geophysical Data Center
(NGDC)\footnotemark. All the GLEs are associated with metric type II
bursts \citep{Gopal10,Gopal11a}, and in many cases, they are reported by more than one
observatories.  

\footnotetext{\url{ftp://ftp.ngdc.noaa.gov/STP/SOLAR\_DATA/SOLAR\_RADIO/SPECTRAL/Type
\_II/Type\_II\_1994-2009}.}

\begin{figure*}[!h]
  \includegraphics[width=1.0\textwidth]{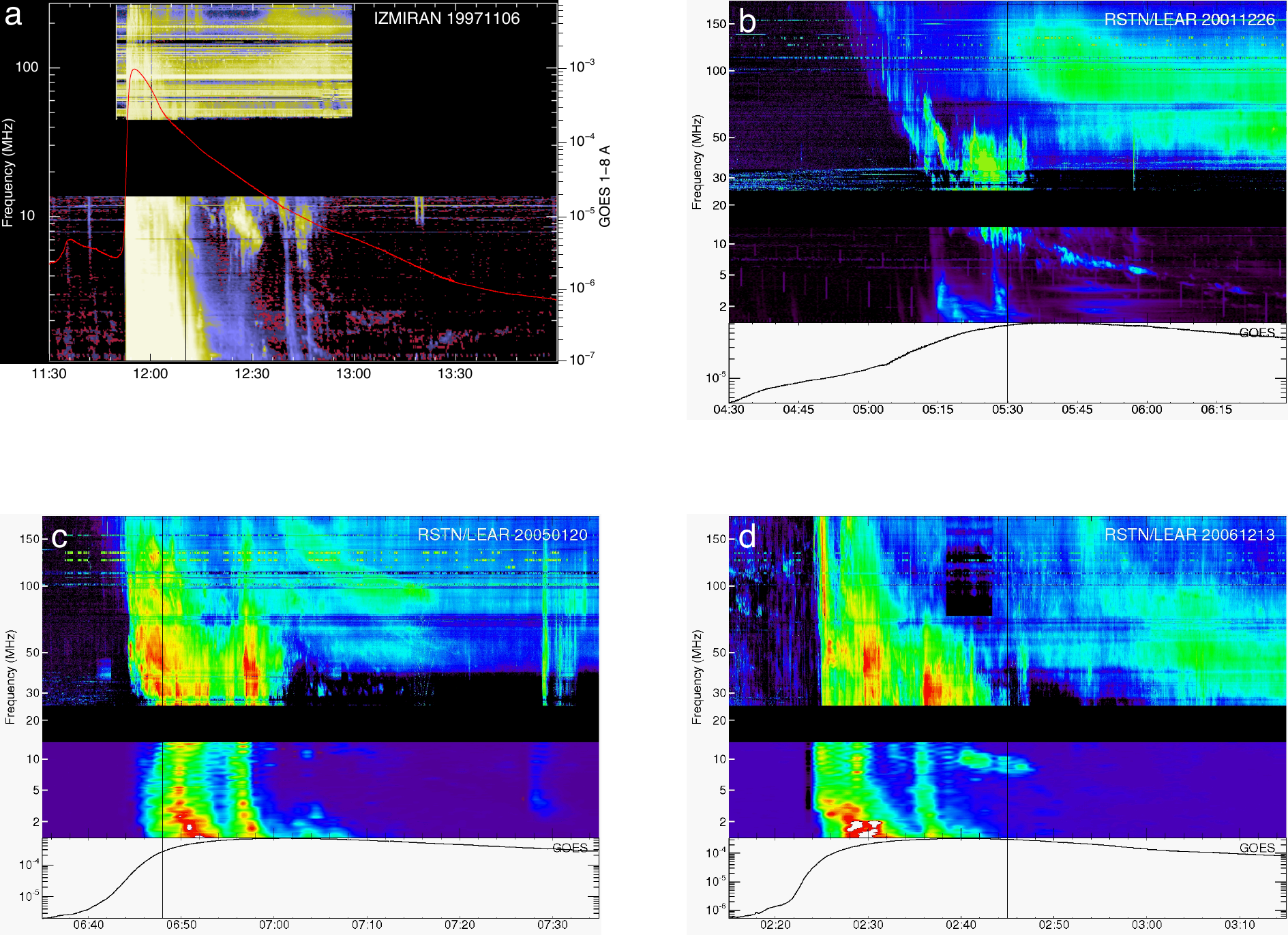}
\caption{Radio dynamic spectra of the same four GLEs as Figure~8 
with soft X-ray 
light curves of the associated flares (courtesy of the
  Living with a Star Coordinated Data Analysis Workshop in 2002 for
  (a) and of S. White for (b)\,--\,(d).)  The vertical line in each
  panel indicates the observed GLE onset.}
\label{fig:9}       
\end{figure*}

However, we do not find it straightforward to isolate metric type II bursts 
in radio dynamic spectra 
around the times of GLEs \citep[see][]{White12}.  
In Figure~9, 
we present them for the same GLEs shown in Figures~5 and 8.  
It appears that only panel (b) gives a 
type II burst clearly from the metric to decametric-hectometetric 
(DH) ranges \citep{Cliver04}.  For GLE~70 shown in panel (d), two
observatories reported a type II burst to continue to 02:44~UT. 
But after 02:40~UT a slow drift is seen below 14 MHz, and its metric
counterpart is not unambiguously found at earlier times.  Other
features are also present in the metric range.  Therefore we
consider the type II burst in GLE~70 to be marginal.  In GLEs~55 and 69
(panels (a) and (c)) the spectra are too complex to allow discrete
type II bursts to be identified, suggesting multiple shocks or other
processes. This may be related to the large-scale disturbances that 
typically did not appear as EIT waves.  Their presence is not clear
in GLE~63, which instead shows the metric type II burst more clearly than
others.  

Although such large-scale disturbances may play a role 
in the detection of SEPs with quick onsets from
poorly-connected regions such as GLE 65 (whose associated global
dimming is discussed by \citet{Mandrini07}), examples such as GLEs 55
and 69 are well-connected events, in which the acceleration
region or shock does not have to move large distances before particles
are injected.  Therefore, we fail to establish the possible importance
of coronal or EIT waves in GLEs.  A possibility still exists, however,
that these large-scale disturbances may bear information on when and
where CME-driven shocks form, and become effective in accelerating
particles.
It has been shown that 
the CMEs reach the height of $\approx$2~$R_{\sun}$ when the shocks
form and accelerate particles
\citep{Reames09a, Reames09b}; see slightly different estimates 
of the CME height by, 
for example, \citet{Kahler94} and \citet{Gopal11a}. 
The relevance of the large-scale disturbances to CME-driven shocks 
needs to be investigated in future GLEs using coronal images with
high cadence \citep[e.g.,][]{Gopal11b,Gopal11c}
in comparison with the CME development in coronagraph
images.  In the past CME shocks have been observed in coronagraph images 
as a streamer deflection
or a diffuse region around the CME \citep{Vourlidas03,Ontiveros09}, 
typically further from the Sun than the height
range corresponding to GLE onsets. 

Having discussed the difficulty to isolate metric type II bursts in
dynamic spectra and to detect coronal waves in EUV and X-ray images
for GLEs in the past, we believe that CME-driven shocks are the main
ingredient of GLEs.  All the GLEs in Solar Cycle 23 were accompanied
by DH type II bursts \citep{Gopal11a}.

\section{Discussion}
\label{sec:5}
The initial task given to us was to find properties of
ARs that unmistakably distinguish GLEs.  ARs have not been widely 
discussed in the context of SEP events, 
and GLEs in particular \citep{Nitta03_icrc,Gopal05_iau,Kahler11b}.  
From the outset it was not clear
if such an attempt could lead to meaningful results. 
First, the question could be ill-posed unless we know 
how special GLEs are as compared with
other SEP events which also manifest in, for example, $>$100~MeV
protons.  In fact, \cite{Li11} have not distinguished GLEs from SEP events.
Similar views of unclear boundaries between GLEs and SEP events 
are expressed by others (e.g., I. Richardson,
private communication 2011).
GLEs are registered when secondary particles
are detected by ground-based detectors,
and it is not straightforward to
define GLEs quantitatively in terms of proton flux at certain
energies.  Direct measurement of protons from space may 
greatly alleviate this difficulty.
However, we are not aware of future plans to measure GeV protons from
space.  Instead the next-generation {\it GOES} satellites will have
differential proton channels only up to 500 MeV\footnotemark, 
down from 700~MeV for
the past and present {\it GOES} satellites.  
Together with the historical events, this
makes ground-based neutron monitors more valuable.

\footnotetext{See \url{http://www.goes-r.gov/spacesegment/seiss.html}.}

Second, it is likely CME-driven shock waves that directly produce 
high-energy SEP events \citep[e.g.,][]{Reames99, Kahler11b}.   Therefore
CMEs and their associated flares may contain more direct information on
the shock properties than do the source ARs.  So we 
study the general properties of  
CMEs and flares.  Consistent with many studies
\citep[e.g.,][]{Gopal10,Gopal11a}, 
we confirm that GLEs tend to be associated with intense 
flares and energetic CMEs.  But it is difficult to set thresholds 
that unconditionally distinguish GLEs.  Furthermore, we should
remember a few GLEs before Solar Cycle 23 that were not associated
with major flares \citep{Cliver06}

Then we at last study the properties of ARs that may be used to
isolate GLEs.  One of the most straightforward measures is the age. 
Old ARs were responsible for GLEs in earlier cycles \citep{Svestka69}
and SEPs in general \citep{Belov05}.  However, the GLEs in Solar
Cycle~23, with one exception, appeared to come from ARs less than 
one rotation old. The sample size is yet too small to discuss 
the possible solar cycle dependence of the origin of GLEs.
Next we confirm that the ARs associated with GLEs tend to be 
magnetically complex, as repeatedly shown in the literature for 
flare-productive ARs \citep[e.g.,][and references cited therein]{Leka03}.  
The total unsigned
magnetic flux both in the whole AR and near the polarity 
inversion line \citep[e.g.][]{Schrijver07} is 
higher for GLEs than for other complex ARs.
In addition, compared with ARs producing SEP events but
not GLEs, GLE ARs tend to be compact.  We, however, need to remember that
these findings are compromised by various observational
restrictions. In order to conduct a statistically study, we only extract
the properties of ARs when they were not far from disk
center, i.e., not capturing the
properties right before the events.  The analysis also has
large error bars and is subject to a small sample size 
(especially for GLEs).
We have to go beyond the basic properties as presented here 
to determine if the ARs associated with GLEs are really special.   
We also need
techniques to handle ARs close to the limb with vector data
\citep[e.g.,][]{Wang09}, so that more useful information on the AR
around the time of the GLE can be
extracted.  For some flare-productive ARs, 
time variations of vector field have been 
studied and non-linear force free field extrapolations have been conducted
\citep[e.g.,][]{Leka03,DeRosa09}, but they are still too few to
distinguish GLE ARs from others. 

The source ARs of GLEs tend to be located in the western hemisphere, 
which is frequently synonymous with 
good magnetic field connection to Earth
in the context of
SEP events including GLEs \citep[see, for example,][]{Kahler11b,Gopal11a}.
In particular, the flare associated with the most intense GLE, 
namely GLE 69, occurred around W60, which is often thought to be 
central in well-connected regions.
This leads us to study magnetic field connection of the GLE ARs and
surrounding regions, using the PFSS model rather than assuming that
the Parker spiral goes straight down to the photosphere.  This is now
routinely done for impulsive SEP or electron events 
\citep{Wang06,Nitta06,Rust08}. 
Using the same technique, we find that a
half of the GLE ARs are well-connected.
We may expect magnetic field connection to influence the onset temporal
behavior of SEP events.  Indeed we can find a few well-connected GLEs
that had a quick onset with respect to the associated flare and CME.
However, the GLE onset time is not strongly correlated with
how well-connected the AR is.  Furthermore,
we may also anticipate more flare-accelerated particles
from well-connected events as manifesting, for example, 
in elevated Fe/O
ratios \citep{Cane03,Cane06,Mewaldt11}.  
However, apart from a couple of well-connected GLEs with
elevated Fe/O ratios, our study indicates that there is hardly a
correlation of the Fe/O with magnetic connection. 

Following \citet{Cliver82b}, we expect injection of particles in a SEP
event to begin when the acceleration region first intersects 
the open field lines connecting to Earth.  For the 1 September 1971
GLE, the proton injection nearly coincided with the time an imaged
type II burst at 22\,--\,55~MHz swept over the footpoint of the
nominal Parker spiral connected to Earth \citep{Cliver82a}.  
Therefore it is important
to locate the acceleration region with respect to well-connected field
lines.  Unfortunately, this could not be done for our GLEs.
CME-driven shocks have been observed by LASCO 
\citep{Vourlidas03,Ontiveros09}, but they are typically further 
from the Sun than the height range corresponding to GLE onsets.  
This may partly reflect 
LASCO's $\gtrsim$20~minute cadence and 
lack of coverage below 2~$R_{\sun}$ from
disk center.  The inner coronagraph (COR1) of 
the Sun Earth Connection Coronal and Heliospheric Investigation
\citep[SECCHI;][]{Howard08} instrument suite on 
the Solar TErrestrial RElations Observatory \citep[STEREO;][]{Kaiser08}
can largely improve this situation for high-energy SEP events 
during the rising phase of
Solar Cycle 24.  The stereoscopic view of the origin and propagation
of CMEs is extremely useful.  
This is exemplified by \cite{Rouillard11}, who 
nicely explained the long delay of a recent non-GLE SEP event 
with respect to the associated CME, 
using observations
at multiple locations and over large distances in combination with
numerical simulations.   

In a majority of GLEs, EUV images from the EIT 
indicate that an extended area far from the flaring AR 
is disturbed during the impulsive phase of the associated flare.  
We may imagine this to be caused by coronal waves.  
However, unlike near-relativistic electron events delayed 
with respect to type III bursts \citep{Krucker99,Haggerty02}, GLEs are
seldom associated with clear EIT waves.  This can be due to the
12~minute cadence of the EIT, which may leave fast 
(e.g, $\gtrsim$1000~km/s) waves undetected.
Around the time we find large-scale
disturbances, a type II burst is also reported, 
even though the actual radio dynamic spectrum often appears to be
so complex that isolation of the type II burst is not trivial
\citep{White12}. GLE~63, in which we do not clearly see the
large-scale disturbances, instead shows a type II burst from the
metric to DH ranges.  If the type II burst 
is a manifestation of the
nose of the CME-driven shock, it is possible that the hypothetical fast wave 
may contribute to the complex dynamic spectrum.  Such a wave may
result from flare reconnection or represent part of the CME-driven
shock away from the nose.  Although we could not make a case where
a wave like this has to do with a GLE, it is possible that it may
correspond to the flank of the CME-driven shock.  
There we expect efficient particle acceleration, because at the flank
quasi-perpendicular conditions may be more naturally met 
than close to the nose.

This is a mere speculation at this stage.  
But we hope to explore the relationship of the CME-driven shock and
large-scale disturbances including EIT waves (which need to be
redefined), comparing high-cadence images from the Atmospheric Imaging
Assembly \citep[AIA;][]{Lemen12} on the 
{\it Solar Dynamics Observatory} (SDO)
with coronagraph images from 
COR1 and COR2 on STEREO.  GLE particles are released when the CME is
at the height of (2\,--\,4)$R_{\sun}$ 
\citep{Reames09a,Reames09b},but this does not
necessarily equate with the height of the shock responsible for particle
acceleration.  
AIA has already observed large-scale
disturbances as fast as $\approx$1000~km/s, but as of the second half
of 2011 
no major gradual SEP events, not to mention GLEs, have occurred yet.

With a few channels that include lines forming at high temperatures, 
the AIA can also observe flare-associated ejections, typically
detected in soft X-rays \citep{Shibata95}.  These ejections have been 
one of the observables used to advocate the ``standard'' 2-d reconnection
model for flares \citep[e.g.,][]{McKenzie02}, although some
observational problems of the prototype event have been pointed out
\citep{Nitta10}.  In the SEP context, the high Fe/O ratios in gradual SEP
events have been shown to be associated with ejections that start
without a pre-acceleration phase \citep{Nitta03}.  
The flares that produce these ejections tend to be impulsive and
spatially not as extended as others associated, for example, with GLEs 59
and 62 (which have low Fe/O ratios). 
GLEs 55\,--\,57 and 60 belong to this category, but we could not
study the possible correlation of flare-associated ejections with the
SEP compositions in events after 2001, when {\it Yohkoh}
observations were terminated.  AIA data will let us characterize
flare-associated ejections in more detail, especially in terms of
their relevance to CME-driven shocks and the large-scale disturbances, 
and eventually to the variable SEP compositions.

\begin{acknowledgements}
We thank the referees for constructive comments, which have greatly
improved the manuscript. 
This work is supported by NASA grants NNX08AB23G, NNX07AN13G and
NNG05GK05G.  We are thankful to H. Moraal for providing neutron
monitor data, and to M. Shea and D. Smart for enlightening discussions
on how these data should be used and interpreted.
The authors
acknowledge NASA's Living With a Star program to make possible the Coordinated
Data Analysis Workshops on GLEs, without which it would have been much
more difficult to analyze relevant data.  We used the CME catalog that
is generated and maintained at the CDAW Data Center by NASA 
and The Catholic University of America in cooperation with the 
Naval Research Laboratory. SOHO is a project of international 
cooperation between ESA and NASA.  
\end{acknowledgements}



\end{document}